\documentclass[runningheads]{llncs}
\usepackage{graphicx}
\usepackage{xcolor}
\usepackage{amsmath}
\usepackage{mathtools}

\DeclareMathOperator*{\argmax}{arg\,max}

\begin{document}
\title{SAM2CLIP2SAM: Vision Language Model for Segmentation of 3D CT Scans for Covid-19 Detection}

\titlerunning{SAM2CLIP2SAM}
%
\author{
Dimitrios Kollias\inst{1} \and
Anastasios Arsenos\inst{2} \and
James Wingate \inst{3} \and
Stefanos Kollias \inst{2}}
\authorrunning{D. Kollias et al.}
\institute{School of Electronic Engineering \& Computer Science, Queen Mary University of London, United Kingdom \\
\email{d.kollias@qmul.ac.uk}\\
\and
School of Electrical \& Computer Engineering, National Technical University Athens, Greece\\
\email{stefanos@cs.ntua.gr, anarsenos@gmail.com}
\\
\and
Univerity of Lincoln,  School of Computer Science Lincoln, United Kingdom\\
\email{jwingate@lincoln.ac.uk}
}
\maketitle              

\begin{abstract}
This paper presents a new approach for effective segmentation of images that can be integrated into any model and methodology; the paradigm that we choose is classification of medical images (3-D chest CT scans) for Covid-19 detection. Our approach includes a combination of vision-language models that segment the CT scans, which are then fed to a deep neural architecture, named RACNet, for Covid-19 detection. In particular, a  novel framework, named SAM2CLIP2SAM, is introduced for segmentation that leverages the strengths of both Segment Anything Model (SAM) and Contrastive Language-Image Pre-Training (CLIP) to accurately segment the right and left lungs in CT scans, subsequently feeding these segmented outputs into RACNet for classification of COVID-19 and non-COVID-19 cases. 
At first, SAM produces multiple part-based segmentation masks for each slice in the CT scan; then CLIP selects only the masks that are associated with the regions of interest (ROIs), i.e., the right and left lungs; finally SAM is given these ROIs as prompts and generates the final segmentation mask for the lungs.
Experiments are presented across two Covid-19 annotated databases which illustrate the improved performance obtained when our method has been used for segmentation of the CT scans.

\end{abstract}
\section{Introduction}

The application of Deep Learning (DL) techniques in medical image analysis has revolutionized the field, leading to substantial improvements in the accuracy and reliability of diagnoses \cite{miah2024can,tagaris2017assessment,tagaris2018machine,kollias2023btdnet,gerogiannis2024covid,chowdhury2023covidetector}. In pathology and radiology, DL models have proven superior in extracting clinically relevant information from medical images compared to traditional manual assessments, which are often subjective and inconsistent. However, integrating these AI-based methods into routine clinical workflows requires significant development and rigorous validation.

This paper focuses on the diagnosis of COVID-19 using advanced medical image analysis techniques, particularly utilizing three-dimensional (3-D) chest CT scans. 
Advanced approaches target the segmentation and automatic detection of pneumonia regions in the lungs, followed by identifying anomalies associated with COVID-19, such as ground-glass opacities, consolidation, and interlobular septal thickening, which are typically found under the pleura. These methods necessitate large, annotated datasets for effective model training.

To meet this need, the COV19-CT-DB \cite{arsenos2022large} was developed, a comprehensive database of 3-D chest CT scans, which includes 7,756 scans (1,661 annotated as COVID-19  and 6,095 as non-COVID-19 cases) of around 2.5 million CT slices.
In addition, we introduce RACNet, an innovative deep neural network architecture designed to address the challenges of analyzing 3-D CT scans. RACNet is engineered to: i) process volumetric CT scan data, ii) handle the variability in slice numbers across scans, and iii) deliver high diagnostic performance for COVID-19. This architecture incorporates dynamic routing and feature alignment mechanisms that selectively process relevant Recurrent Neural Network (RNN) outputs for making diagnostic decisions.

However the performance of RACNet (and of any classification model) depends on the input 3-D CT scans and whether they are segmented or not.
Segmentation of chest CT scans is crucial for several reasons in the context of classification, particularly for diseases like COVID-19. Firstly, segmentation isolates specific regions of interest (ROIs), such as the lungs, which are critical for diagnosing respiratory conditions. By focusing on these relevant areas, the classification model can avoid distractions from irrelevant regions, leading to more accurate diagnoses. Secondly, segmentation allows for the extraction of detailed and localized features from the segmented regions. This detailed feature extraction can significantly improve the performance of classification models by providing more precise and relevant information. Thirdly, medical images often contain noise and artifacts from surrounding tissues and organs. Segmentation helps in reducing this noise by isolating the lung regions, thus providing cleaner input data for the classification model. Finally, segmentation ensures that the classification model consistently analyzes the same anatomical regions across different scans and patients. This consistency is vital for training robust and generalizable models.

Many state-of-the-art approaches for COVID-19 classification either do not perform segmentation at all or perform segmentation as a combination of morphological transforms  and a pre-trained model, like U-Net  \cite{ref14}. More specifically, for each CT-scan, every slice first passes through the pre-trained U-Net. After all slices of the CT-scan are segmented by the U-Net model, there is a checking procedure to assure that all slices are segmented. If a slice has a mask area less than 40 \% of the largest mask area of the CT-scan, then morphological transforms are used to segment this slice. However, the issue with such approaches is that the segmentation is not very accurate; for instance, the segmentation mask includes the lungs and the mediastinal mass between them, or a lung is segmented along with its surrounding area and so on. To solve this issue, 
in this work, we present an innovative framework that integrates Segment Anything Model (SAM) \cite{sam1} and Contrastive Language-Image Pre-Training (CLIP) \cite{clip1} to segment only the right and left lungs in CT scans, with these outputs subsequently being fed into RACNet for classification to detect COVID-19 and non-COVID-19 cases.


SAM  and CLIP are two exemplary Vision Foundation Models (VFMs) that have showcased exceptional capabilities in segmentation and zero-shot recognition, respectively. SAM, a prompt-driven segmentation model, excels across diverse domains. SAM has been trained on an extensive dataset of over one billion masks, making it highly adaptable to a wide range of downstream tasks through interactive prompts. It can operate in two distinct modes: segment everything mode and promptable segmentation mode. In our approach, we employ both modes to achieve optimal segmentation results. SAM has shown impressive results in a broad range of tasks for natural images, but its performance has not been state-of-the-art when being directly applied to medical imaging. Conversely, CLIP's training with millions of text-image pairs has endowed it with an unprecedented ability in zero-shot visual recognition. 

Despite their individual successes, their unified potential for medical image segmentation remains largely unexplored.
Existing methods for adapting SAM to medical imaging often rely on tuning strategies that require extensive data or prior prompts tailored to the specific task, posing significant challenges when data samples are limited \cite{sam2}. 

Medical imaging segmentation tasks exhibit inherent variability based on the specific clinical scenario, complicating the adaptation process. To assign semantic labels to SAM-provided masks, our method involves cropping the original image according to these masks. This set of cropped regions is then processed by CLIP, which retrieves the corresponding crop in a zero-shot manner using visually descriptive sentences related to the lungs \cite{gpt1}, generated via GPT. The retrieved region of interest (ROI) mask is subsequently used for bounding box prompt generation, guiding SAM to deliver the final lung segmentation.





The remainder of this paper is organized as follows: Section 2 reviews the related work we have developed in medical image analysis and COVID-19 diagnosis and introduces the vision-language models. Section 3 outlines the proposed pipeline, including the vision-language models and the RACNet architecture. Section 4 presents the experimental setup and results, whilst Section 5 concludes the paper and suggests future research directions.

\section {Related Work}


SAM is a promptable vision foundation segmentation model designed to segment everything in an image based on different types of prompts, such as bounding boxes and point prompts. It comprises an image encoder, a prompt encoder, and a lightweight mask decoder. A pretrained Vision Transformer (ViT) \cite{vit} is used as the image encoder, transforming the input image into dense features. The prompt encoder processes prompts, which can be sparse or dense, encoding them into a format suitable for mask generation. SAM can operate in two modes: segment everything mode and promptable segmentation mode. The former segments everything in the image using a grid of keypoints as prompts, while the latter segments specific regions based on provided prompts. Our framework utilizes both modes of SAM in conjunction with CLIP.

CLIP is a pre-trained large Vision-Language Model known for its strong generalizability and impressive zero-shot domain adaptation capabilities. It aligns image and text modalities within a shared embedding space, enabling it to perform image classification directly on the target dataset without any fine-tuning. By employing prompt engineering, CLIP can be adapted to various domains, incorporating relevant semantic details related to the specific target task. Our framework leverages CLIP's zero-shot recognition capabilities to identify and retrieve the ROI in CT scans, facilitating accurate lung segmentation.

Various 3-D CNN models have been employed for detecting COVID-19 and differentiating it from common pneumonia (CP) and normal cases using volumetric 3D CT scans \cite{morani2024covid,morani2024detecting}. For instance,  \cite{jaiswal2020classification} utilized a pretrained DenseNet-201 model, which was fine-tuned on CT scan images to classify them into COVID-19 or non-COVID-19 categories. The performance of this model was compared against other pretrained and fine-tuned models (VGG16, ResNet152V2, and Inception-ResNetV2).
\cite{khadidos2020analysis} combined CNNs with RNNs to process CT scan images, successfully distinguishing between COVID-19 and non-COVID-19 cases. Similarly,  \cite{amyar2020multi} proposed a multi-task architecture featuring a shared encoder for 3D CT scans and three branches: a decoder for image reconstruction, a second decoder for COVID lesion segmentation, and a multi-layer perceptron for classification into COVID-19 and non-COVID-19 categories.

Additionally, a weakly supervised deep learning framework was introduced by \cite{ref2} for COVID-19 classification and lesion localization using 3D CT volumes. This framework utilized a pretrained UNet to segment lung regions in each CT slice, which were then fed into a 3D DNN for classification. COVID-19 lesions were localized by combining activation regions from the DNN with connected components through an unsupervised method.
Furthermore, \cite{paperAAAI} established baseline performance using 3D models such as ResNet3D101 and DenseNet3D121. They then proposed a differentiable neural architecture search (DNAS) framework to automatically identify optimal 3D DL models for CT scan classification. The study also published the training, validation, and test datasets used, providing a resource for future research.

\section{The Proposed Approach}

\begin{figure}[h!]
\centering
\includegraphics[width=1.\linewidth]{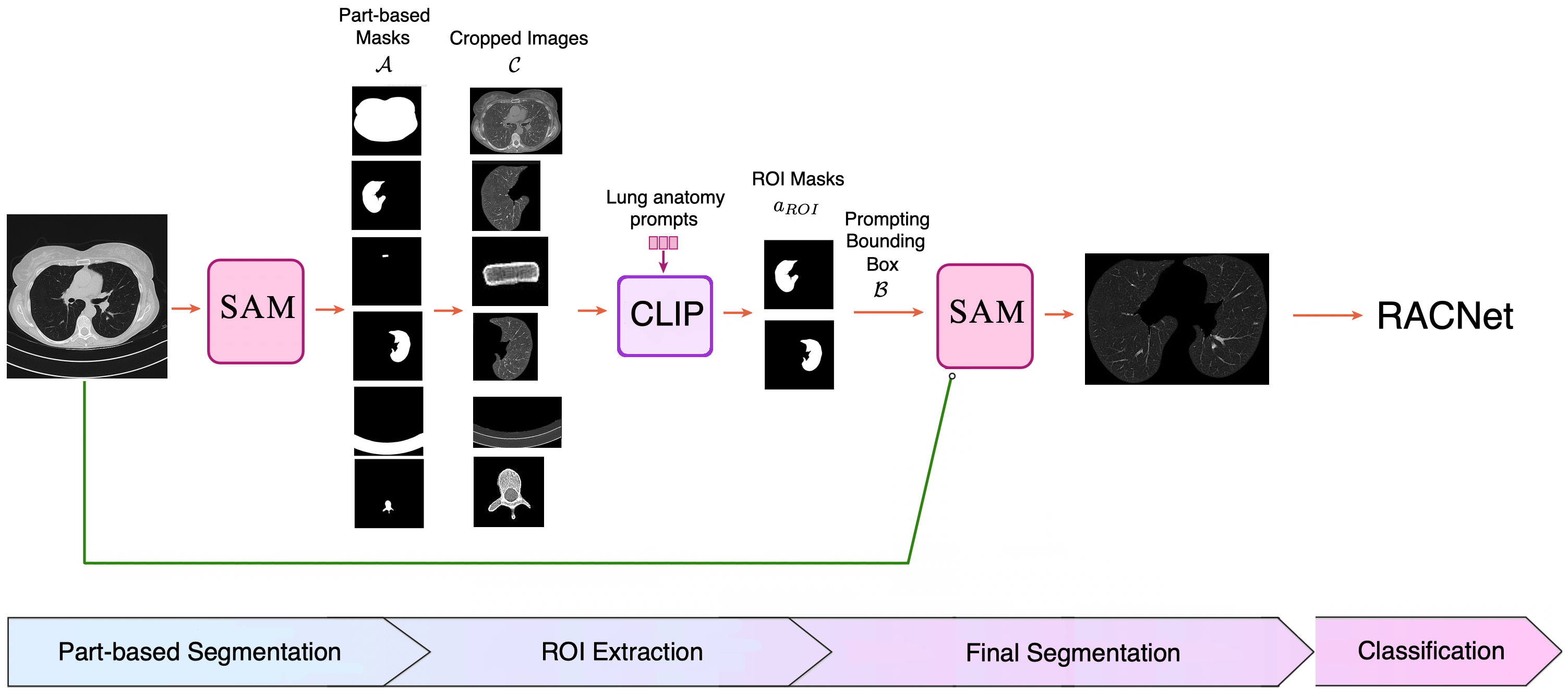}
\caption{Our whole proposed pipeline that includes segmentation and classification tasks}
\label{meth}
\end{figure}

The complete framework of the proposed approach is depicted in Figure \ref{meth}. It consists of the segmentation framework (which has the goal of segmenting only the right and left lungs in CT scans) and the COVID-19 classification framework (RACNet). Let us stress that the proposed segmentation framework can be integrated in any classification model for any similar task (not just RACNet, or not for COVID-19 detection).
More details about each framework follow.

\subsection {The SAM2CLIP2SAM Segmentation Framework}

Our proposed framework leverages the combined strengths of the Segment Anything Model (SAM) and Contrastive Language-Image Pre-Training (CLIP) for the task of lung segmentation in CT scans, followed by classification using RACNet.
This proposed methodology can be delineated into three primary components: i) part-based segmentation using SAM; ii) region of interest (ROI) extraction using CLIP; and iii) final segmentation using bounding box prompts from SAM.

\subsubsection{Part-based Segmentation using SAM}

In the first phase, SAM is employed to generate part-based segmentation masks from the input CT scan image. SAM operates in its segment-everything mode to produce an exhaustive set of masks.
Given an input image I $\in \Re^{H \times W \times 3}$ (with H and W denoting the Height and Width of the image) and a grid of keypoints $\mathcal{G}$, SAM generates a set of part-based masks $\mathcal{A}$:   

\begin{equation}
    \mathcal{A} = SAM_{seg-everything} (I,\mathcal{G})
\end{equation}

\noindent where $\mathcal{A}$ = \{$a_1, a_2, ... ,  a_n$\} and each $a_i$ is a mask corresponding to a specific region in $I$.


\subsubsection{ROI Extraction using CLIP}

Once the part-based masks are generated, the next step involves extracting the mask corresponding to the Region Of Interest (ROI) using CLIP. The masks from $\mathcal{A}$ 
  are used to crop the input image $I$, resulting in a set of cropped images $\mathcal{C}$: 

  \begin{equation}
      \mathcal{C} = \{I \cdot a_i \mid  a_i \in \mathcal{A}, \text{ area($a_i$)}> \tau\}
  \end{equation}

\noindent where $\cdot$ denotes element-wise multiplication and $\tau$ is the area threshold used to filter out background masks.

Visually descriptive textual prompts for lung anatomy are generated using GPT. These prompts are passed through CLIP's text encoder to obtain a textual embedding $\mathcal{W}$:

\begin{equation}
    \mathcal{W} = CLIP_{text-encoder} (VDT)
\end{equation}

\noindent where VDT represents the visually descriptive text. 

Each cropped image  $c \in \mathcal{C}$ is then passed through CLIP's vision encoder to obtain a set of vision embeddings $\mathcal{V}$:

 \begin{equation}
      \mathcal{V} = \{ v_i \mid  v_i =  CLIP_{vision-encoder}(c_i), c_i \in \mathcal{C} \}
  \end{equation}

The similarity between each vision embedding $v_i$ and the text embedding $\mathcal{W}$ is computed using cosine similarity:

\begin{equation}
sim(v_i,\mathcal{W}) =\dfrac{ v_i \cdot \mathcal{W}}{\vert \vert v_i \vert \cdot \vert \vert \mathcal{W} \vert \vert}  
\end{equation}

The mask corresponding to the ROI is identified as the mask with the highest similarity score:

\begin{equation}
a_{ROI} = a_{\argmax_i sim(v_i,\mathcal{W})}  
\end{equation}

\subsubsection{Final Segmentation using Bounding Box Prompts}

The final segmentation step involves using the bounding box of the ROI mask to generate prompts for SAM. The bounding box $\mathcal{B}$ of the ROI mask $a_{ROI}$ is calculated as: 

\begin{equation}
\mathcal{B} = \big ( min_{(x,y) \in a_{ROI}} x, \text{ } min_{(x,y) \in a_{ROI}} y,  \text{ } max_{(x,y) \in a_{ROI}} x, \text{ } max_{(x,y) \in a_{ROI}} y \big )
\end{equation}

SAM is then prompted with $\mathcal{B}$ to generate the final segmentation mask for the lungs:

\begin{equation}
    a_{final} = SAM_{promptable} (I, \mathcal{B})
\end{equation}

The segmented lung regions $a_{final}$ are subsequently used to train and classify images in RACNet for detecting COVID-19 and non-COVID-19 cases.


\subsection{The RACNet COVID-19 Classification Framework}

\begin{figure}[h!]
\centering
\includegraphics[width=1.\linewidth]{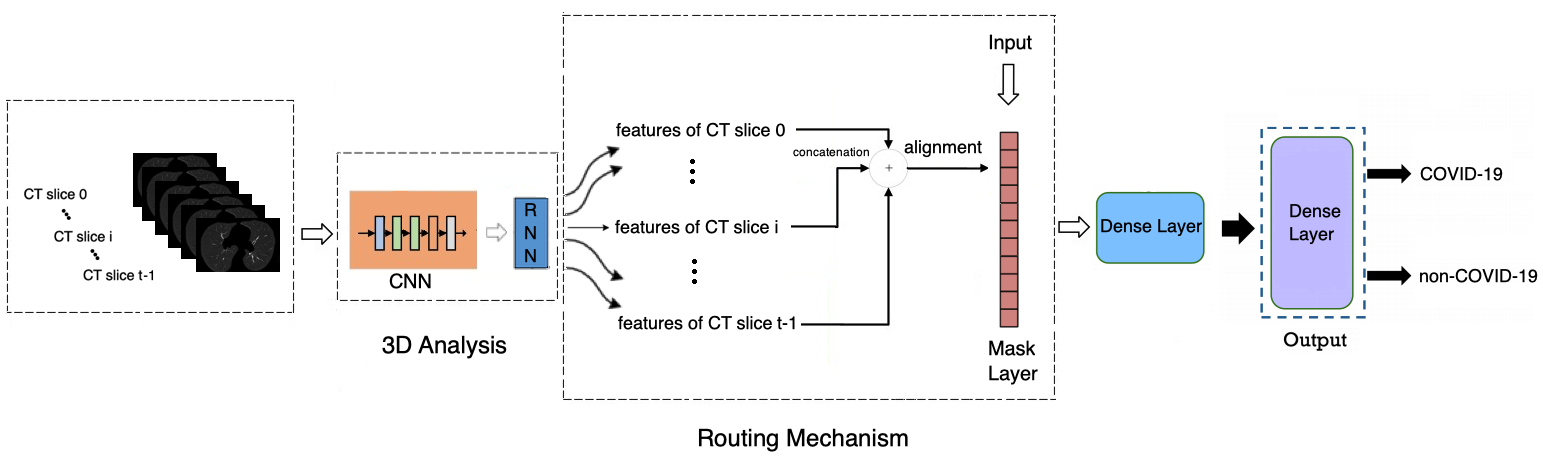}
\caption{The RACNet model for COVID-19 Classification}
\label{methodology_1}
\end{figure}

An overview of the RACNet COVID-19 Classification Framework can be seen in Figure \ref{methodology_1}.
Each segmented 3-D scan, consisting of $t$ slices, is analyzed using a CNN-RNN based model. A routing component equipped with alignment and masking operations effectively handles the variability in $t$ across different CT scans. The final diagnosis is produced through a dense layer followed by an output layer.

The input data is fed into the 3D Analysis component of RACNet, which comprises a Convolutional Neural Network (CNN) and a Recurrent Neural Network (RNN). The CNN component performs 2-D slice analysis, extracting lung features from each slice. Our objective is to replicate the diagnostic process of medical professionals by utilizing all CT slices for COVID-19 detection. The RNN component processes the sequential features extracted by the CNN, analyzing the CT scan slices in sequence.
The features extracted by the RNN are subsequently fed into the RACNet Routing Component. These features are concatenated and then processed by the Mask Layer.

During the training of RACNet, the routing component dynamically selects a number of RNN outputs equal to the length $l$ of the input, zeroing out the remaining RNN outputs. The selected outputs are then fed into the dense layer. This selection process can be implemented in two ways: by either selecting the first $l$ RNN outputs, or through an `alignment' approach, where $l$ RNN outputs are positioned at equidistant intervals within the $[0, t-1]$ range, with the remaining outputs placed between them.

The concatenated features are then processed by the RACNet Classification Component, which includes a dense layer and an output layer utilizing a softmax activation function to predict the presence or absence of COVID-19.

The dense layer is responsible for the final extraction of semantic information from the RNN outputs. During training, weight updating is guided by the routing mechanism and the Mask Layer. Only the weights corresponding to the selected RNN outputs are updated, while the others remain constant until they are selected for another scan input. We keep these weights constant during training and ignore the links of RNN outputs which are not selected by the routing mechanism \cite{b300}. 

\section{Experimental study}

\subsection{Databases}

The COV19-CT-DB database comprises 7,756 3-D chest CT scans collected from various medical institutions. Each CT scan contains between 50 and 700 2-D CT slices. The dataset includes 1,661 scans from COVID-19 positive patients and 6,095 from non-COVID-19 cases, totaling approximately 2.5 million images. All images have been anonymized, with 724,273 slices labeled as COVID-19 and 1,775,727 slices labeled as non-COVID-19. There is big variability in scan lengths due to factors such as required resolution and the specific features of the imaging equipment used \cite{10193437}. To extend and disseminate this research, four COV19D Competitions have been organized in conjunction with workshops at ICCV 2021 \cite{ref12}, ECCV 2022 \cite{kollias2022ai}, ICASSP 2023 \cite{10193422} and CVPR 2024 \cite{kollias2024domain}. These competitions featured challenges on COVID-19 detection, severity assessment, and domain adaptation \cite{kollias2017adaptation,ref10}, all leveraging the COV19-CT-DB database. In the presented results thereafter, we utilized the part of COV19-CT-DB \cite{arsenos2022large,ref12} used in the ECCV 2022 Competition \cite{kollias2022ai,Tailor,springer}. This includes 1550 COVID-19 and 5044 non-COVID-19 3-D CT scans. The training part includes 882 COVID-19 and 1110 non-COVID-19 samples and its validation set  215 COVID-19 and 289 non-COVID-19 samples. 

We also used the MosMedData database \cite{ref13}, which includes   856 COVID-19 CT-scans and 254 non-COVID samples (601 COVID-19 and 178 non-COVID-19 in the training set; 256  COVID-19 and 76 non-COVID-19 in the test set).

\subsection{Performance Metric}

The performance measure used to evaluate the models' performance in detecting Covid-19 is the average F1 Score  (i.e., macro F1 Score) across all 2 categories (Covid-19 and non-Covid-19) :

\begin{equation} \label{expr}
\mathcal{P} = \frac{ F_1^{Covid-19} + F_1^{non-Covid-19}}{2}
\end{equation}

The $F_1$ score is a weighted average of the recall (i.e., the ability of the classifier to find all the positive samples) and precision (i.e., the ability of the classifier not to label as positive a sample that is negative). The $F_1$ score  takes values in the range $[0,1]$; high values are desired. The $F_1$ score is defined as:

\begin{equation} \label{f1}
F_1 = \frac{2 \times precision \times recall}{precision + recall}
\end{equation}

\subsection{Implementation Details}

We employed ViT-H, a variant of SAM, and ViT-L/14 trained in CLIP by OpenAI. The visually descriptive textual sentences were generated using GPT-3.5. All was implemented in PyTorch. All experiments were conducted on  Tesla V100 32GB GPU.

Regarding implementation of RACNet: i) we used EfficientNetB0 as CNN model, stacking a global average pooling layer on top, a batch normalization layer and dropout (with keep probability 0.8) \cite{kollias2018deep}; ii) we used a single one-directional GRU layer consisting of 128 units as RNN model; iii) the first dense layer consisted of 128 hidden units. 
Regarding implementation details of RACNet training, batch size was equal to 5 (i.e, at each iteration our model processed 5 CT scans) and the input length 't' (see Figure \ref{methodology_1}) was 700 (the maximum number of slices found across all CT scans).
Our model was fed with 3-D CT scans composed of CT slices; each slice was resized from its original size of $512 \times 512$ to  $256 \times 256$. Loss function was cross entropy. Adam optimizer was used with learning rate $10^{-4}$. 


\subsection{Experimental Results}

In the following we provide an extensive experimental study comparing various networks and segmentation methods on the two above described databases.

Figure \ref{db1} shows the performance comparison (in terms of  $F_1$ Score) of RACNet (trained and tested on COV19D-CT-DB  of the ECCV 2022 Competition) when its input is: i) unsegmented images; ii)  segmented images with the conventional approach \cite{salpea}; iii) segmented images with our proposed SAM2CLIP2SAM framework. It is evident that when segmentation is performed with our SAM2CLIP2SAM framework, RACNet achieves 3.75\% and 1.75\% superior performance to the cases when no segmentation is performed, or segmentation is performed with the conventional approach, respectively. This validates our notion that our segmentation approach enhances the classification model's feature extraction and assists it into focusing on only the important ROIs which are the left and right lungs. This also validates our observation that conventional segmentation approaches contain some noise and artifacts, whereas our method removes them.

\begin{table}
\caption{Performance comparison on the test set of  COV19D-CT-DB (of the ECCV 2022 Competition) between RACNet and the state-of-the-art, when images are segmented with the conventional approach and when images are segmented with our proposed  SAM2CLIP2SAM framework. F1 Score is given in \% }
\label{db1}
\centering
\scalebox{1.1}{
\begin{tabular}{ |c||c|c|c| }
 \hline
 \multicolumn{1}{|c||}{\begin{tabular}{@{}c@{}} \begin{tabular}{@{}c@{}} COV19D-CT-DB (ECCV 2022) \end{tabular} \end{tabular}} & 
\multicolumn{3}{c|}{\begin{tabular}{@{}c@{}}  F1   \end{tabular}}
\\ 
  \hline
   & Macro & COVID & non-COVID \\
 \hline
 \hline
   MDAP \cite{MDAP_ch_1}&   \begin{tabular}{@{}c@{}}  87.87    \end{tabular} & \begin{tabular}{@{}c@{}} 78.80    \end{tabular} &
\begin{tabular}{@{}c@{}} 96.95   \end{tabular} 
\\ 
\hline

   MDAP with SAM2CLIP2SAM&   \begin{tabular}{@{}c@{}}  89.87    \end{tabular} & \begin{tabular}{@{}c@{}} 81.50    \end{tabular} &
\begin{tabular}{@{}c@{}} 97.25   \end{tabular} 
\\ 
\hline

 FDVTS \cite{FDVTS_ch_1}&   \begin{tabular}{@{}c@{}}  89.11 \end{tabular} & \begin{tabular}{@{}c@{}}  80.92 \end{tabular} &
\begin{tabular}{@{}c@{}}  97.31    \end{tabular}  \\ 
\hline

 FDVTS with SAM2CLIP2SAM&   \begin{tabular}{@{}c@{}}  90.61 \end{tabular} & \begin{tabular}{@{}c@{}}  82.22 \end{tabular} &
\begin{tabular}{@{}c@{}}  97.51    \end{tabular}  \\ 
\hline

 ACVLab \cite{ACVLab_ch_1}&   \begin{tabular}{@{}c@{}}  89.11    \end{tabular} & \begin{tabular}{@{}c@{}} 80.78    \end{tabular} &
\begin{tabular}{@{}c@{}} 97.45   \end{tabular} 
\\ 
\hline

 ACVLab with SAM2CLIP2SAM&   \begin{tabular}{@{}c@{}}  90.61    \end{tabular} & \begin{tabular}{@{}c@{}} 82.08    \end{tabular} &
\begin{tabular}{@{}c@{}} 97.65   \end{tabular} 
\\ 
\hline

 RACNet without segmentation &  
\begin{tabular}{@{}c@{}}   93.06    \end{tabular} 
& \begin{tabular}{@{}c@{}}   92.18    \end{tabular} 
& \begin{tabular}{@{}c@{}}   93.95   \end{tabular}
\\ 
\hline

 RACNet with conventional segmentation &  
\begin{tabular}{@{}c@{}}   95.06    \end{tabular} 
& \begin{tabular}{@{}c@{}}   94.18    \end{tabular} 
& \begin{tabular}{@{}c@{}}   95.95   \end{tabular}
\\ 
\hline

 \textbf{RACNet with SAM2CLIP2SAM} &  
\begin{tabular}{@{}c@{}}   \textbf{96.81}    \end{tabular} 
& \begin{tabular}{@{}c@{}}   \textbf{95.68}    \end{tabular} 
& \begin{tabular}{@{}c@{}}   \textbf{97.95}   \end{tabular}
\\ 
\hline
\end{tabular}
}
\end{table}

\begin{figure}[h!]
\centering
\includegraphics[height=3cm]{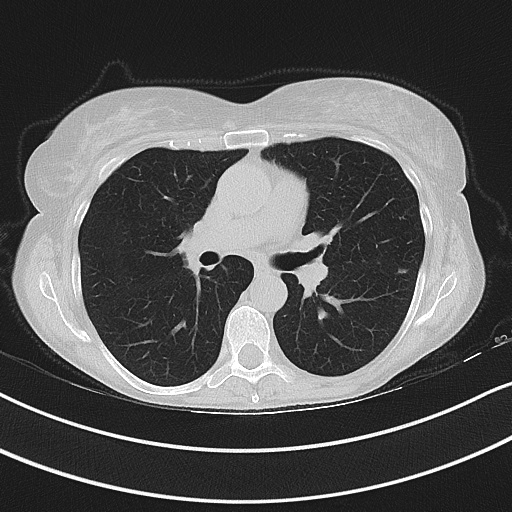} 
\includegraphics[height=3cm]{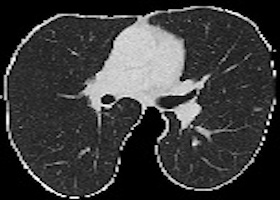} 
\includegraphics[height=3cm]{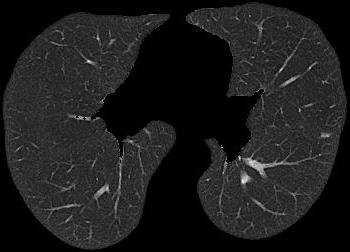} \\
\includegraphics[height=3cm]{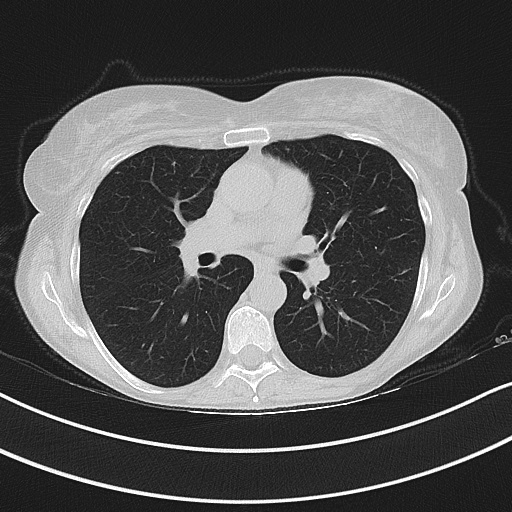} 
\includegraphics[height=3cm]{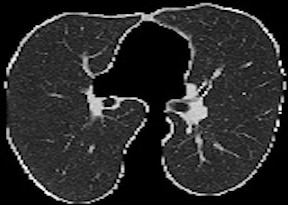} 
\includegraphics[height=3cm]{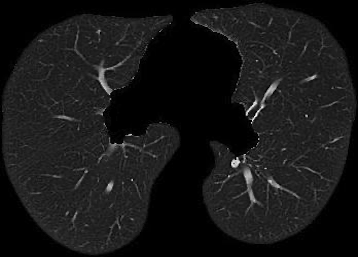} \\ 
\includegraphics[height=3cm]{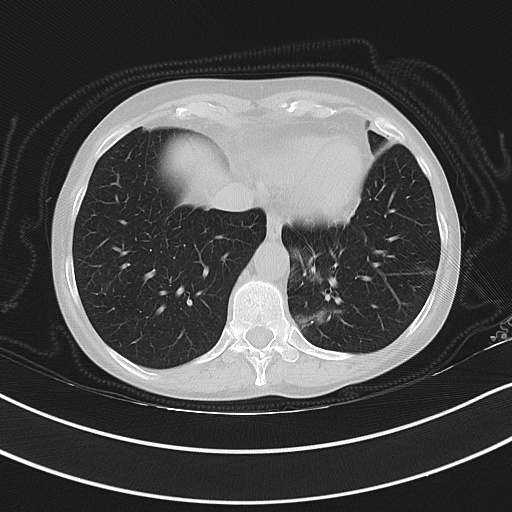} 
\includegraphics[height=3cm]{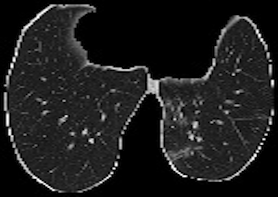} 
\includegraphics[height=3cm]{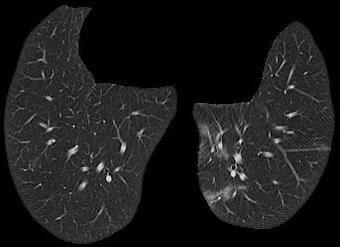} \\
 Original-Unsegmented \hspace{0.25cm} Conventionally segmented\hspace{0.15cm} SAM2CLIP2SAM segmented
\caption{Illustration of the improvement in segmentation quality when the CT scan slices are segmented with our proposed approach, the SAM2CLIP2SAM framework (right column) vs when they are segmented with conventional approaches (middle column).  
}
\label{CCC}
\end{figure}

Figure \ref{CCC} illustrates this further. It presents three cases of unsegmented slices of a CT scan (left column), along with their cases when they are segmented with conventional approaches (middle column) and with our proposed framework SAM2CLIP2SAM (right column). It is evident that the segmentation result with our approach is more accurate and error-prune. In the first case (top row), the mediastinal mass between the left and right lungs is kept when the slices are segmented with conventional approaches, whereas it is not kept (i.e., it is black) when the slices are segmented with our SAM2CLIP2SAM framework. 
In all cases, one can also note that a bit of the pleural space (e.g. on the peripheral of the lungs) is also kept and is not masked when the slices are segmented with conventional approaches; this is not the case when the slices are segmented with our SAM2CLIP2SAM framework.

Additionally,  Figure \ref{db1} presents a comparison between the performance of RACNet to that of the state-of-the-art in the COV19D-CT-DB of the ECCV 2022 Competition. One can see that RACNet, especially when trained with CT scans segmented with our proposed SAM2CLIP2SAM framework, outperforms (in terms of F1 Score) all  state-of-the-art methods by large margins (between 7.69\% and 8.94\%). Finally, Figure \ref{db1} shows that our proposed SAM2CLIP2SAM framework can be applied to the state-of-the-art methods as well, enhancing their performance by between 1.5\% and 2\% (most improvement in performance is seen for Covid-19 class, which is the most important).

\begin{table}[h]
\caption{Performance comparison on the test set of  MosMedData between RACNet and the state-of-the-art, when images are segmented with the conventional approach and when images are segmented with our proposed  SAM2CLIP2SAM framework. F1 Score, Precision and Sensitivity are given in \% }
\label{db2}
\centering
\scalebox{1.1}{
\begin{tabular}{|c|c|c|c|}
\hline
  MosMedData   & Precision & Sensitivity & F1 \\
 \hline
\hline

   MC3\_18 \cite{MC318}&  79.43 & 98.43 & 87.92  \\
\hline

   MC3\_18 with SAM2CLIP2SAM &  80.93 & 98.73 & 88.95  \\
\hline

 Densenet3D121 \cite{Densenet} &   84.23 & 92.16 & 88.01 \\
\hline

 Densenet3D121 with SAM2CLIP2SAM &   85.73 & 93.66 & 89.51 \\
\hline

 CovidNet3D \cite{paperAAAI} &   79.50 & 98.82 & 88.11
\\
\hline

 CovidNet3D with SAM2CLIP2SAM &   81.50 & 98.92 & 89.37
\\
\hline

 EMARS-APS  \cite{MICCAIEMARS}&   93.52 & 90.59& 92.03  \\
\hline

 EMARS-APS with SAM2CLIP2SAM&   95.02 & 92.09& 93.53  \\
\hline

 RACNet without segmentation &   92.69 & 90.85& 91.76 \\
\hline

 RACNet with conventional segmentation &   94.69 &92.85& 93.76 \\
\hline

\textbf{RACNet with SAM2CLIP2SAM}  &\textbf{96.12}&\textbf{94.86}& \textbf{95.49} \\
\hline

\end{tabular}
}
\end{table}

Table \ref{db2} shows the performance comparison (in terms of  $F_1$ Score) of RACNet (trained and tested on MosMedData) when its input is: i) unsegmented images; ii)  segmented images with the conventional approach \cite{salpea}; iii) segmented images with our proposed SAM2CLIP2SAM framework. It is evident that when segmentation is performed with our SAM2CLIP2SAM framework, RACNet achieves 3.43\% and 1.43\% superior performance to the cases when no segmentation is performed, or segmentation is performed with the conventional approach, respectively. Additionally,  Figure \ref{db2} presents a comparison between the performance of RACNet to that of the state-of-the-art on MosMedData. One can see that RACNet, especially when trained with CT scans segmented with our proposed SAM2CLIP2SAM framework, outperforms (in terms of F1 Score) all  state-of-the-art methods by large margins (between 16.69\% and 2.6\%). Finally, Figure \ref{db1} shows that our proposed SAM2CLIP2SAM framework can be applied to the state-of-the-art methods as well, enhancing their performance by between 1.03\% and 2\%.



\section{Conclusions and Future Work}

In this paper, we introduced a novel approach for COVID-19 diagnosis that leverages the RACNet deep neural architecture combined with vision-language models. Our method utilizes the SAM and CLIP models to perform precise segmentation of the right and left lungs in CT scans. The segmented outputs are then fed into RACNet, which demonstrates enhanced accuracy in detecting COVID-19 and non-COVID-19 cases.

Our experimental results highlight the significant benefits of integrating vision-language models with deep learning architectures for medical image analysis. The proposed method shows promising potential for managing the variability of medical images collected from different healthcare institutions. Furthermore, this research defines a critical direction for future studies, including the exploration of uncertainty estimation \cite{arsenos2024uncertainty} to further improve diagnostic performance \cite{arsenos2024common,karampinis2024ensuring,arsenos4674579nefeli}.

The findings of this study suggest that the combination of advanced segmentation techniques and robust classification models can substantially enhance the accuracy and reliability of COVID-19 detection. Future work will focus on refining these techniques and exploring their applicability to other medical imaging tasks, with the aim of creating more efficient and scalable diagnostic frameworks.

%
%

%
%
\bibliographystyle{splncs04}
\bibliography{bibtex}

%
%
%
%




\end{document}